# Quantum dialogue without information leakage based on the entanglement swapping between any two Bell states and the shared secret Bell state

Tian-Yu Ye*, Li-Zhen Jiang

College of Information & Electronic Engineering, Zhejiang Gongshang University, Hangzhou 310018, P.R.China

**Abstract**

In order to avoid the risk of information leakage during the information mutual transmission between two authorized participants, i.e., Alice and Bob, a quantum dialogue protocol based on the entanglement swapping between any two Bell states and the shared secret Bell state is proposed. The proposed protocol integrates the ideas of block transmission, two-step transmission and unitary operation encoding together using the Bell states as the information carriers. Besides the entanglement swapping between any two Bell states, a shared secret Bell state is also used to overcome the information leakage problem, which not only makes Bob aware of the prepared initial state but also is used for Bob's encoding and entanglement swapping. Security analysis shows that the proposed protocol can resist the general active attacks from an outside eavesdropper Eve. Moreover, the relation between the maximal amount of information Eve can gain and the detection probability is derived.

## 1 Introduction

Different from the classical cryptography, quantum cryptography is based on the physical law of quantum mechanics rather than the mathematical complexity. Quantum cryptography can be attributed into several types, such as quantum key distribution (QKD), quantum secure direct communication (QSDC), quantum secret sharing (QSS) and so on. The aim of QKD is to establish a shared random key between two remote authorized users. Since Bennett and Brassard put forward the first QKD protocol (BB84)[1] in 1984, many other theoretical QKD protocols have been put forward[2-6]. Different from QKD, QSDC allows secret messages to be transmitted directly without creating a key to encrypt them in advance. In 2002, Long and Liu[4] proposed the first QSDC. In this protocol, two ordered particle sequences derived from the order Bell states are transmitted in blocks from Alice to Bob in two steps with two security checks. It should be emphasized that Long and Liu's protocol is also the first protocol employing the ideas of block transmission and two-step transmission. Since then, the block transmission technique has been widely adopted. For example, in 2003, Deng et al. [7] also presented a two-step QSDC based on block transmission and two-step transmission. This protocol encodes information by applying unitary operations on Bell states. In other words, this protocol adopts the unitary operation encoding. Until now, a lot other QSDC protocols have been proposed with different implementation ways so that QSDC has gained considerable development [8-17]. In 2002, Bostrom and Felbinger[8] proposed the famous Ping-Pong protocol using a Bell state. Later, Cai and Li[9] improved the capacity of Ping-Pong protocol by introducing two additional unitary operations. In 2004, Deng and Long[10] put forward an one-time pad QSDC. In 2005, Wang et al.[11] proposed a multi-step QSDC using the multi-particle Green-Horne-Zeilinger state. In 2007, Li et al.[12] presented a QSDC with quantum encryption based on the pure entangled state. In 2008, Chen et al.[13] presented a controlled QSDC protocol with quantum encryption using a partially entangled GHZ state; Chen et al.[14] also presented a three-party controlled QSDC protocol based on the W state. In 2011, Yang et al.[15] proposed two fault tolerant two-step QSDC protocols which are robust against the collective-dephasing noises and the collective-rotation noises. In 2012, Li et al.[16] improved the eavesdropping detection efficiency of Deng et al.'s two-step QSDC[7] by using the four-particle cluster state; Song and Wang[17] reviewed the recent development in quantum communication including QKD, QSDC, QSS, quantum teleportation and so on. They also pointed out that multi-step transmission, block transmission and order rearrangement are popularly used to construct quantum communication protocols, and that single photons and entangled photon pairs are often used as the information carriers. With regard to the relation between QKD and QSDC, it should be emphasized that all QSDC protocols are QKD protocols at the same time, but not all QKD protocols are QSDC protocols, such as the BB84 QKD protocol [1].

Unfortunately, most of those existing QSDC protocols [4,7-16] belong to the kind of message-unilaterally-transmitted communication protocol, where two authorized users can not exchange their individual secret messages simultaneously. In fact, the bidirectional and multi-directional simultaneous communications are very useful in practice. Fortunately, in 2004, Zhang et al.[18-19] and Nguyen[20] proposed the new concept of quantum dialogue, which greatly improved the above drawback. Subsequently, Man et al. [21] thought that Nguyen' protocol[20] cannot resist the intercept-and-resend attack, and put forward an method to solve it. Later, a lot novel quantum dialogue protocols of different types have been put forward, such us the protocols based on a GHZ state[22-23], based on entanglement swapping of GHZ states[24], based on a single photon[25-26], based on entanglement swapping of Bell sates[27],and based on entanglement swapping of two Bell states via cavity QED[28]. However, during the earliest development of quantum dialogue, researchers were not aware of the problem of information leakage so that it happens in lots of earliest quantum dialogue protocols [18-28]. In 2007, Man and Xia[29] began to recognize the information definite leakage problem in Jin's protocol[22], and suggested a solution towards it. Later, in 2008, Gao et al. [30-31] discovered the information leakage problem in Jin et al.'s protocol[22], Man and Xia's improved version[29], Nguyen's protocol[20], Man et al. 's protocol[21], Ji and Zhang's protocol[25] and Man et al.'s protocol[24], and analyzed it in detail from the viewpoint of information theory and cryptography; Tan and Cai[32] analyzed the information leakage problem of Nguyen' protocol[20] from the viewpoint of Holevo Bound. Afterwards, researchers began to concentrate on solving the information leakage problem in quantum dialogue. As a result, some information leakage resistant quantum dialogue protocols were derived, such as the protocols based on a shared private quantum state[33-34], based on the correlation extractability of Bell state and the auxiliary particle[35], based on the measurement correlation from entanglement swapping between two Bell states[36]. In addition, Ye and Jiang[37] suggested two methods to solve the information definite leakage problem in Man and Xia's protocol [23]in 2013. However, their two protocols[37]still run the risk of information leakage.[38-39]

---

*Corresponding author:
 E-mail：happyyty@aliyun.com

In this paper, we integrate the ideas of block transmission [4,7], two-step transmission[4,7] and unitary operation encoding[7] together to construct a novel quantum dialogue protocol with the Bell states being the information carriers. Besides the entanglement swapping between any two Bell states, a shared secret Bell state is also used to overcome the information leakage problem. The shared secret Bell state has two functions in the dialogue process: firstly, it makes Bob aware of the prepared initial state; secondly, it is used for Bob's encoding and entanglement swapping. Security analysis shows that the proposed protocol can resist the general active attacks from an outside eavesdropper Eve. Moreover, the relation between the maximal amount of information Eve can gain and the detection probability is derived.

## 2 Quantum dialogue protocol

The four Bell states can be expressed as follows:

$$|\Phi^+\rangle = \frac{1}{\sqrt{2}}(|00\rangle + |11\rangle) = \frac{1}{\sqrt{2}}(|+\rangle|+\rangle + |-\rangle|-\rangle) \tag{1}$$

$$|\Phi^-\rangle = \frac{1}{\sqrt{2}}(|00\rangle - |11\rangle) = \frac{1}{\sqrt{2}}(|+\rangle|-\rangle + |-\rangle|+\rangle) \tag{2}$$

$$|\Psi^+\rangle = \frac{1}{\sqrt{2}}(|01\rangle + |10\rangle) = \frac{1}{\sqrt{2}}(|+\rangle|+\rangle - |-\rangle|-\rangle) \tag{3}$$

$$|\Psi^-\rangle = \frac{1}{\sqrt{2}}(|01\rangle - |10\rangle) = \frac{1}{\sqrt{2}}(|+\rangle|-\rangle - |-\rangle|+\rangle) \tag{4}$$

Here, $|\pm\rangle = (|0\rangle \pm |1\rangle)/\sqrt{2}$. It is well known that $I = |0\rangle\langle 0| + |1\rangle\langle 1|$, $\sigma_x = |0\rangle\langle 1| + |1\rangle\langle 0|$, $i\sigma_y = |0\rangle\langle 1| - |1\rangle\langle 0|$ and $\sigma_z = |0\rangle\langle 0| - |1\rangle\langle 1|$ are four unitary operations. Here, both Alice and Bob agree on in advance that each unitary operation corresponds to a two-bit secret message in the way like $I \to 00$, $\sigma_x \to 01$, $i\sigma_y \to 10$ and $\sigma_z \to 11$.

It is easy to verify that the entanglement swapping between $|\Psi^-\rangle$ and arbitrary one of the four Bell states can be described by formulas (5-8), where the subscripts $A_i$ and $B_i$ ($i = 1, 2$) denote the two particles from a Bell state, respectively.

$$|\Psi^-\rangle_{A_1B_1} \otimes |\Psi^-\rangle_{A_2B_2} = \frac{1}{2}\left\{|\Phi^+\rangle_{A_1A_2}|\Phi^+\rangle_{B_1B_2} - |\Phi^-\rangle_{A_1A_2}|\Phi^-\rangle_{B_1B_2} - |\Psi^+\rangle_{A_1A_2}|\Psi^+\rangle_{B_1B_2} + |\Psi^-\rangle_{A_1A_2}|\Psi^-\rangle_{B_1B_2}\right\} \tag{5}$$

$$|\Psi^-\rangle_{A_1B_1} \otimes |\Psi^+\rangle_{A_2B_2} = \frac{1}{2}\left\{|\Phi^-\rangle_{A_1A_2}|\Phi^+\rangle_{B_1B_2} - |\Phi^+\rangle_{A_1A_2}|\Phi^-\rangle_{B_1B_2} - |\Psi^-\rangle_{A_1A_2}|\Psi^+\rangle_{B_1B_2} + |\Psi^+\rangle_{A_1A_2}|\Psi^+\rangle_{B_1B_2}\right\} \tag{6}$$

$$|\Psi^-\rangle_{A_1B_1} \otimes |\Phi^-\rangle_{A_2B_2} = \frac{1}{2}\left\{|\Phi^+\rangle_{A_1A_2}|\Psi^+\rangle_{B_1B_2} - |\Phi^-\rangle_{A_1A_2}|\Psi^-\rangle_{B_1B_2} - |\Psi^+\rangle_{A_1A_2}|\Phi^+\rangle_{B_1B_2} + |\Psi^-\rangle_{A_1A_2}|\Phi^-\rangle_{B_1B_2}\right\} \tag{7}$$

$$|\Psi^-\rangle_{A_1B_1} \otimes |\Phi^+\rangle_{A_2B_2} = \frac{1}{2}\left\{|\Phi^-\rangle_{A_1A_2}|\Psi^+\rangle_{B_1B_2} - |\Phi^+\rangle_{A_1A_2}|\Psi^-\rangle_{B_1B_2} + |\Psi^-\rangle_{A_1A_2}|\Phi^+\rangle_{B_1B_2} - |\Psi^+\rangle_{A_1A_2}|\Phi^-\rangle_{B_1B_2}\right\} \tag{8}$$

According to formulas (5-8), if the Bell-basis measurements are performed on $A_1A_2$ and $B_1B_2$, respectively, each initial state group of $A_1B_1$ and $A_2B_2$ will collapse to four different outcome combinations of $A_1A_2$ and $B_1B_2$ with the same probability after entanglement swapping. Moreover, each outcome combination of $A_1A_2$ and $B_1B_2$ solely corresponds to one initial state group of $A_1B_1$ and $A_2B_2$. Corresponding to formulas (5-8), four collections composed by different outcome combinations of $A_1A_2$ and $B_1B_2$ are labeled as:

$$\left\{ |\Phi^+\rangle_{A_1A_2}|\Phi^+\rangle_{B_1B_2}, |\Phi^-\rangle_{A_1A_2}|\Phi^-\rangle_{B_1B_2}, |\Psi^+\rangle_{A_1A_2}|\Psi^+\rangle_{B_1B_2}, |\Psi^-\rangle_{A_1A_2}|\Psi^-\rangle_{B_1B_2} \right\} \to \mathbb{C}_0 \tag{9}$$

$$\left\{ |\Phi^-\rangle_{A_1A_2}|\Phi^+\rangle_{B_1B_2}, |\Phi^+\rangle_{A_1A_2}|\Phi^-\rangle_{B_1B_2}, |\Psi^+\rangle_{A_1A_2}|\Psi^-\rangle_{B_1B_2}, |\Psi^-\rangle_{A_1A_2}|\Psi^+\rangle_{B_1B_2} \right\} \to \mathbb{C}_1 \tag{10}$$

$$\left\{ |\Phi^+\rangle_{A_1A_2}|\Psi^+\rangle_{B_1B_2}, |\Phi^-\rangle_{A_1A_2}|\Psi^-\rangle_{B_1B_2}, |\Psi^+\rangle_{A_1A_2}|\Phi^+\rangle_{B_1B_2}, |\Psi^-\rangle_{A_1A_2}|\Phi^-\rangle_{B_1B_2} \right\} \to \mathbb{C}_2 \tag{11}$$

$$\left\{ |\Phi^-\rangle_{A_1A_2}|\Psi^+\rangle_{B_1B_2}, |\Phi^+\rangle_{A_1A_2}|\Psi^-\rangle_{B_1B_2}, |\Psi^-\rangle_{A_1A_2}|\Phi^+\rangle_{B_1B_2}, |\Psi^+\rangle_{A_1A_2}|\Phi^-\rangle_{B_1B_2} \right\} \to \mathbb{C}_3 \tag{12}$$

Further extending the first Bell state in each initial state group from $|\Psi^-\rangle$ to other three Bell states, all the collections composed by different outcome combinations of $A_1A_2$ and $B_1B_2$ after entanglement swapping can be summarized as Table1.

Table 1. The outcome collections of entanglement swapping between any two Bell states

|  | $|\Phi^+\rangle_{A_2B_2}$ | $|\Phi^-\rangle_{A_2B_2}$ | $|\Psi^+\rangle_{A_2B_2}$ | $|\Psi^-\rangle_{A_2B_2}$ |
|---|---|---|---|---|
| $|\Phi^+\rangle_{A_1B_1}$ | $\mathbb{C}_0$ | $\mathbb{C}_1$ | $\mathbb{C}_2$ | $\mathbb{C}_3$ |
| $|\Phi^-\rangle_{A_1B_1}$ | $\mathbb{C}_1$ | $\mathbb{C}_0$ | $\mathbb{C}_3$ | $\mathbb{C}_2$ |
| $|\Psi^+\rangle_{A_1B_1}$ | $\mathbb{C}_2$ | $\mathbb{C}_3$ | $\mathbb{C}_0$ | $\mathbb{C}_1$ |
| $|\Psi^-\rangle_{A_1B_1}$ | $\mathbb{C}_3$ | $\mathbb{C}_2$ | $\mathbb{C}_1$ | $\mathbb{C}_0$ |

Suppose that Alice has a secret consisting of $2N$ bits, i.e., $\{(i_1,j_1)(i_2,j_2)\cdots(i_n,j_n)\cdots(i_N,j_N)\}$, and Bob has a secret consisting of $2N$ bits, i.e., $\{(k_1,l_1)(k_2,l_2)\cdots(k_n,l_n)\cdots(k_N,l_N)\}$, where $i_n,j_n,k_n,l_n \in \{0,1\}$, $n \in \{1,2,\cdots,N\}$. We integrate the ideas of block transmission [4,7], two-step transmission[4,7] and unitary operation encoding[7] together to construct a quantum dialogue protocol with the Bell states being the information carriers. That is to say, the two ordered particle sequences from the order Bell states are transmitted from Alice to Bob in two steps here. In the meanwhile, the secret is encoded with unitary operations performed on the ordered particle sequences.

**Step1: Preparation.** Alice prepares $2N$ Bell states $\{(A_1,B_1),(A_2,B_2),\cdots,(A_{2N},B_{2N})\}$ in order with each two adjacent Bell states $(\{(A_{2n-1},B_{2n-1}),(A_{2n},B_{2n})\}, n=1,2,\cdots,N)$ in the same state. Here, the subscript denotes the order of each Bell state. She extracts all the particles $A$ from these Bell states in order to form the sequence $S_A$. Likewise, the remaining particles $B$ form the sequence $S_B$. That is, $S_A = \{A_1,A_2,\cdots,A_{2N}\}$ and $S_B = \{B_1,B_2,\cdots,B_{2N}\}$. For the first security check, Alice prepares another batch of sample Bell states, and randomly inserts the sample particles $A$ and $B$ into the sequences $S_A$ and $S_B$, respectively. As a result, $S_A$ and $S_B$ turn into two new sequences $S_A'$ and $S_B'$, respectively. Then, Alice sends $S_B'$ to Bob, and keeps $S_A'$ in her hand. After Bob confirms Alice that he has received $S_B'$, they start the first security check to test whether $S_B'$ was eavesdropped during the transmission.

**Step2: Alice's encoding.** After dropping out the sample particles, $S_A'$ and $S_B'$ turn back into $S_A$ and $S_B$, respectively. Both Alice and Bob divide their own sequences into groups (a group contains two adjacent particles). That is, $(A_{2n-1},A_{2n})$ forms the group $n$ in $S_A$, while $(B_{2n-1},B_{2n})$ forms the group $n$ in $S_B$. They agree on that the particle $A$ performed with Alice's unitary operation in each group should have a different position with the particle $B$ performed with Bob's unitary operation in the corresponding group. Concretely speaking, in group $n$, once Alice performs her unitary operation $U^A_{i_n j_n}$ on the particle $A_{2n-1}$ $(A_{2n})$, Bob should perform his unitary operation $U^B_{k_n l_n}$ on the particle $B_{2n}$ $(B_{2n-1})$. Without loss of generality, assume that Alice performs $U^A_{i_n j_n}$ on $A_{2n-1}$ for encoding her two-bit secret $(i_n,j_n)$ here. For the second security check, Alice prepares a large number of single sample particles, which is randomly in one of the four states $\{|0\rangle,|1\rangle,|+\rangle,|-\rangle\}$, and randomly inserts them into $S_A$. As a result, $S_A$ turns into a new sequence $S_A''$. Then, Alice sends $S_A''$ to Bob. After Bob confirms Alice that he has received $S_A''$, they start the second security check to test whether $S_A''$ was eavesdropped during the transmission.

**Step3: Quantum dialogue.** After dropping out the single sample particles, $S_A^{''}$ turns back into $S_A$ again. Having both $S_A$ and $S_B$ in his hand, Bob picks up one particle from each sequence in order, and stores each two adjacent Bell states as a group. That is, group $n$ contains two adjacent Bell states $\{(U_{i_n j_n}^A A_{2n-1}, B_{2n-1}), (A_{2n}, B_{2n})\}$. For simplicity, we just deal with the group 1 to illustrate the dialogue process (other groups are similar). In group 1, $\{(U_{i_1 j_1}^A A_1, B_1), (A_2, B_2)\}$, Bob firstly performs the Bell-basis measurement on $(A_2, B_2)$. Consequently, Bob can know the initial state of group 1 prepared by Alice in Step 1. According to his Bell-basis measurement outcome, Bob reproduces a new $(A_2, B_2)$ with no state measurement performed. Then Bob encodes his two-bit secret $(k_1, l_1)$ by performing $U_{k_1 l_1}^B$ on the new particle $B_2$. As a result, group 1 turns into $\{(U_{i_1 j_1}^A A_1, B_1), (A_2, U_{k_1 l_1}^B B_2)\}$. Afterward, Bob performs the Bell-basis measurements on $(U_{i_1 j_1}^A A_1, A_2)$ and $(B_1, U_{k_1 l_1}^B B_2)$, respectively. According to formulas (9-12), Bob can know which outcome collection $\{(U_{i_1 j_1}^A A_1, A_2), (B_1, U_{k_1 l_1}^B B_2)\}$ belongs to. Then, Bob announces Alice this outcome collection through classical channel. After hearing of Bob's announcement, Alice can infer out the four possible kinds of states about $\{(U_{i_1 j_1}^A A_1, B_1), (A_2, U_{k_1 l_1}^B B_2)\}$ from Table 1. Furthermore, since Alice prepares the initial state herself, according to her own unitary operation $U_{i_1 j_1}^A$, Alice is able to read out Bob's two-bit secret $(k_1, l_1)$ from Table 1. Similarly, according to the outcome collection $\{(U_{i_1 j_1}^A A_1, A_2), (B_1, U_{k_1 l_1}^B B_2)\}$ belongs to, Bob can also infer out the four possible kinds of states about $\{(U_{i_1 j_1}^A A_1, B_1), (A_2, U_{k_1 l_1}^B B_2)\}$ from Table 1. Since Bob knows the prepared initial state from the Bell-basis measurement on $(A_2, B_2)$, according to his own unitary operation $U_{k_1 l_1}^B$, Bob is able to read out Alice's two-bit secret $(i_1, j_1)$ from Table 1. Therefore, Alice and Bob have accomplished the dialogue process successfully.

Here we also use the first two Bell states $(A_1, B_1)$ and $(A_2, B_2)$ to further explain the dialogue process. Suppose that Alice's first two-bit secret $(i_1, j_1)$ is **01**, and Bob's first two-bit secret $(k_1, l_1)$ is **11**. Moreover, assume that the first two Bell states $(A_1, B_1)$ and $(A_2, B_2)$ are prepared by Alice in the state of $|\Psi^-\rangle_{A_1 B_1}$ and $|\Psi^-\rangle_{A_2 B_2}$, respectively. Consequently, $(A_1, B_1)$ and $(A_2, B_2)$ will change along the following manner

$$\left. \begin{array}{l} |\Psi^-\rangle_{A_1 B_1} \Rightarrow \sigma_x^A \otimes |\Psi^-\rangle_{A_1 B_1} = |\Phi^-\rangle_{A_1 B_1} \\ |\Psi^-\rangle_{A_2 B_2} \Rightarrow \sigma_z^B \otimes |\Psi^-\rangle_{A_2 B_2} = |\Psi^+\rangle_{A_2 B_2} \end{array} \right\} \Rightarrow \left\{ \begin{array}{l} |\Phi^-\rangle_{A_1 A_2} |\Psi^+\rangle_{B_1 B_2} \\ |\Phi^+\rangle_{A_1 A_2} |\Psi^-\rangle_{B_1 B_2} \\ |\Psi^-\rangle_{A_1 A_2} |\Phi^+\rangle_{B_1 B_2} \\ |\Psi^+\rangle_{A_1 A_2} |\Phi^-\rangle_{B_1 B_2} \end{array} \right\} \rightarrow \mathbb{C}_3 \tag{13}$$

The $\mathbb{C}_3$ is published by Bob through classical channel. According to three known information: the outcome collection $\mathbb{C}_3$, the prepared initial state $|\Psi^-\rangle$ and her own unitary operation $\sigma_x^A$, Alice can read out from Table 1 that Bob's two-bit secret $(k_1, l_1)$ is **11**. Likewise, according to the outcome collection $\mathbb{C}_3$, the initial state $|\Psi^-\rangle$ known from his Bell-basis measurement on $(A_2, B_2)$ and his own unitary operation $\sigma_z^B$, Bob can read out from Table 1 that Alice's two-bit secret $(i_1, j_1)$ is **01**. Apparently, during the dialogue process, $(A_2, B_2)$ acts as a shared secret Bell state between Alice and Bob. It has two functions: firstly, it makes Bob aware of the prepared initial state; secondly, it is used for Bob's encoding and entanglement swapping.

## 3 Security checks and security analysis

As it is well known to all, a quantum secure communication protocol should possess good security. Otherwise, it will be meaningless. In the literature of quantum secure communication, the security check is always adopted to guarantee its security.

In the proposed protocol, since Alice sends two particle sequences composed by Bell states to Bob in two steps, two security checks are needed in total. The first is used to check the transmission security of $S_B^{'}$, while the second is to check the transmission security of $S_A^{''}$. The first security check is implemented in the following way: (a) Alice firstly tells Bob the positions of the sample particles $B$ in $S_B^{'}$; (b) Bob chooses randomly one of the two sets of measuring basis, i.e., $Z$-basis $(\{|0\rangle,|1\rangle\})$ and $X$-basis $(\{|+\rangle,|-\rangle\})$, to measure the sample particles $B$, and tells Alice his measurement basis and measurement outcomes; (c) Alice chooses the same measurement basis as Bob's to measure the sample particles $A$ in $S_A^{'}$; (d) Alice judges whether there is an eavesdropping by comparing their measurement outcomes. If there is no eavesdropping, their measurement outcomes should be highly correlated, according to formulas (1-4), and they continue the communication. Otherwise, they halt the communication. The second security check is implemented as follows: (a) Alice firstly tells Bob the positions and the preparation basis of the single sample particles in $S_A^{"}$; (b) Bob measures the single sample particles in the same basis as the preparation basis of Alice and tells Alice his measurement outcomes; (c) Alice judges whether there is an eavesdropping by comparing the initial states of the single sample particles with Bob's measurement outcomes. If there is no eavesdropping, they continue the communication. Otherwise, they halt the communication.

In fact, during the second transmission, an outside eavesdropper Eve can only disturb the transmission of $S_A^{"}$ and can not steal any secret, since no one can distinguish a Bell state by only obtaining one particle. The reason lies in that particle $A$ is a complete mixed state, since its reduced density matrix is $\rho_A = tr_B\{|\Phi^\pm\rangle\langle\Phi^\pm|\} = tr_B\{|\psi^\pm\rangle\langle\psi^\pm|\} = \frac{1}{2}I$. Therefore, the security of the proposed protocol mainly depends on the transmission of $S_B^{'}$. In essence, the first security check uses the entanglement correlation between two particles from a Bell state for eavesdropping detection. This security check method has been popularly used in some previous protocols [7,9,21,33,35,36,37], and its effectiveness has been widely accepted. The detailed analysis of its effectiveness against the general active attacks like the intercept-resend attack, the measure-resend attack and the entangle-measure attack has been made in Refs.[33,35,36,37].

Now let us turn to analyze how much information Eve can maximally gain. First suppose that Alice sends $|0\rangle$ to Bob. Then the state of the system composed of Bob's particle and Eve's probe can be expressed as[7]

$$|\psi^{'}\rangle = \hat{E}|0,\varepsilon\rangle \equiv \hat{E}|0\rangle_B|\varepsilon\rangle = \alpha|0\rangle_B|\varepsilon_{00}\rangle + \beta|1\rangle_B|\varepsilon_{01}\rangle \equiv \alpha|0,\varepsilon_{00}\rangle + \beta|1,\varepsilon_{01}\rangle, \tag{14}$$

$$\rho^{'} = |\alpha|^2|0,\varepsilon_{00}\rangle\langle 0,\varepsilon_{00}| + |\beta|^2|1,\varepsilon_{01}\rangle\langle 1,\varepsilon_{01}| + \alpha\beta^*|0,\varepsilon_{00}\rangle\langle 1,\varepsilon_{01}| + \alpha^*\beta|1,\varepsilon_{01}\rangle\langle 0,\varepsilon_{00}|. \tag{15}$$

where $\hat{E}$ is an attack unitary operation, $|\varepsilon\rangle$ is the state of an ancilla, $|\varepsilon_{a,b}\rangle$ $(a,b \in (0,1))$ are pure ancillary states uniquely determined by $\hat{E}$, and $|\alpha|^2 + |\beta|^2 = 1$. After encoding of the unitary operations $I$, $\sigma_x$, $i\sigma_y$ and $\sigma_z$ with the probabilities $p_0$, $p_1$, $p_2$ and $p_3$, respectively, the state reads[7]

$$\rho^{"} = (p_0+p_3)|\alpha|^2|0,\varepsilon_{00}\rangle\langle 0,\varepsilon_{00}| + (p_0+p_3)|\beta|^2|1,\varepsilon_{01}\rangle\langle 1,\varepsilon_{01}| + (p_0-p_3)\alpha\beta^*|0,\varepsilon_{00}\rangle\langle 1,\varepsilon_{01}| + (p_0-p_3)\alpha^*\beta|1,\varepsilon_{01}\rangle\langle 0,\varepsilon_{00}|$$
$$+ (p_1+p_2)|\alpha|^2|1,\varepsilon_{00}\rangle\langle 1,\varepsilon_{00}| + (p_1+p_2)|\beta|^2|0,\varepsilon_{01}\rangle\langle 0,\varepsilon_{01}| + (p_1-p_2)\alpha\beta^*|1,\varepsilon_{00}\rangle\langle 0,\varepsilon_{01}| + (p_1-p_2)\alpha^*\beta|0,\varepsilon_{01}\rangle\langle 1,\varepsilon_{00}|, \tag{16}$$

which can be rewritten in the orthogonal basis $\{|0,\varepsilon_{00}\rangle, |1,\varepsilon_{01}\rangle, |1,\varepsilon_{00}\rangle, |0,\varepsilon_{01}\rangle\}$,

$$\rho^{"} = \begin{pmatrix} (p_0+p_3)|\alpha|^2 & (p_0-p_3)\alpha\beta^* & 0 & 0 \\ (p_0-p_3)\alpha^*\beta & (p_0+p_3)|\beta|^2 & 0 & 0 \\ 0 & 0 & (p_1+p_2)|\alpha|^2 & (p_1-p_2)\alpha\beta^* \\ 0 & 0 & (p_1-p_2)\alpha^*\beta & (p_1+p_2)|\beta|^2 \end{pmatrix}, \tag{17}$$

where $p_0 + p_1 + p_2 + p_3 = 1$.

The information $I_0$ Eve can gain is equal to the Von Neumann entropy[7]

$$I_0 = \sum_{i=0}^{3} -\lambda_i \log_2 \lambda_i, \tag{18}$$

where $\lambda_i$ $(i=0,1,2,3)$ are the eigenvalues of $\rho^{"}$, and can be expressed as

$$\lambda_{0,1} = \frac{1}{2}(p_0+p_3) \pm \frac{1}{2}\sqrt{(p_0+p_3)^2 - 16p_0p_3|\alpha|^2|\beta|^2} = \frac{1}{2}(p_0+p_3) \pm \frac{1}{2}\sqrt{(p_0+p_3)^2 - 16p_0p_3(d-d^2)}, \tag{19}$$

$$\lambda_{2,3} = \frac{1}{2}(p_1+p_2) \pm \frac{1}{2}\sqrt{(p_1+p_2)^2 - 16p_1p_2|\alpha|^2|\beta|^2} = \frac{1}{2}(p_1+p_2) \pm \frac{1}{2}\sqrt{(p_1+p_2)^2 - 16p_1p_2(d-d^2)}, \tag{20}$$

where $d$ is the detection probability. In the case of $p_0 = p_1 = p_2 = p_3 = \frac{1}{4}$, where Alice encodes exactly 2 bits, we have $\lambda_0 = \frac{1}{2}d$, $\lambda_1 = \frac{1}{2}(1-d)$, $\lambda_2 = \frac{1}{2}d$ and $\lambda_3 = \frac{1}{2}(1-d)$. As a result, the maximal amount of information Eve can gain in this case is equal to the Shannon entropy of a binary channel:

$$I_0(d) = -d\log_2(\frac{1}{2}d) - (1-d)\log_2(\frac{1}{2} - \frac{1}{2}d). \tag{21}$$

Then suppose that Alice sends $|1\rangle$ to Bob. After the above security analysis is done in full analogy, the same crucial relations can be obtained. The maximal amount of information Eve can gain in this case is also equal to the Shannon entropy of a binary channel:

$$I_1(d) = -d\log_2(\frac{1}{2}d) - (1-d)\log_2(\frac{1}{2} - \frac{1}{2}d). \tag{22}$$

Therefore, the maximal amount of information Eve can gain is

$$I(d) = \frac{1}{2}(I_0(d) + I_1(d)) = 1 - d\log_2 d - (1-d)\log_2(1-d). \tag{23}$$

The function $I(d)$ is plotted in Fig.1. Obviously, $I(d)$ reaches its maximum at $d = 1/2$, and can be inversed on the interval $[0, 1/2]$, resulting in a monotonous function, $0 \leq d(I) \leq 1/2$, $I \in [1, 2]$. In order to gain a desired information $I > 1$ per attack, Eve has to encounter a detection probability $d(I) > 0$. If Eve wants to gain the full information $(I = 2)$, the detection probability will be $d(I = 2) = \frac{1}{2}$.

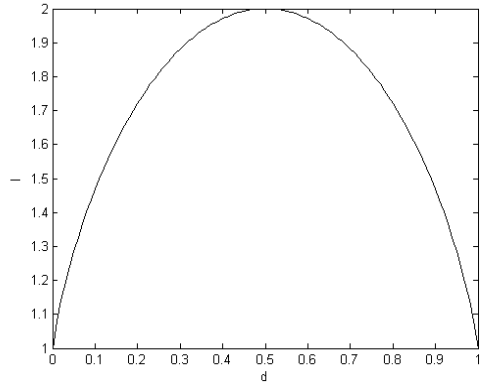

Fig. 1 The relation between the maximal amount of information Eve can gain and the detection probability

## 4 Discussions

**(1) The information leakage problem**

Here we consider the passive attack named as information leakage. We also use the first two Bell states $(A_1, B_1)$ and $(A_2, B_2)$ to analyze it. Although Eve hears from Bob the outcome collection $\{(U^A_{i_1 j_1} A_1, A_2), (B_1, U^B_{k_1 l_1} B_2)\}$ belongs to, she still cannot read out any information about the secret, due to no knowledge about the initial state of the shared secret quantum state $(A_2, B_2)$. Consequently, no information leaks out to Eve. On the other hand, the information leakage problem can also be analyzed via the perspective of information theory. Since Eve is not aware of $(A_2, B_2)$, with respect to the unitary operation combinations performed by Alice and Bob, the outcome collection $\{(U^A_{i_1 j_1} A_1, A_2), (B_1, U^B_{k_1 l_1} B_2)\}$ belongs to contains sixteen kinds of choosing, corresponding to $-\sum_{i=1}^{16} p_i \log_2 p_i = -16 \times \frac{1}{16}\log_2 \frac{1}{16} = 4$ bits for Eve. Fortunately, this amount of information for Eve is equal to the total amount of secret transmitted between Alice and Bob. Hence, no information leakage happens in the proposed protocol. Obviously, besides the entanglement swapping between any two Bell states, the shared secret quantum state $(A_2, B_2)$ also plays a key role in avoiding the information leakage problem.

**(2) The information-theoretical efficiency**

The information-theoretical efficiency defined by Cabello[3] is $\eta = b_s / (q_t + b_t)$, where $b_s$, $q_t$ and $b_t$ are the expected secret bits received, the qubits used and the classical bits exchanged between Alice and Bob. In the proposed protocol, the first two adjacent Bell states $(A_1,B_1)$ and $(A_2,B_2)$ can be used for exchanging four bits in total, i.e., two bits from Alice and two bits from Bob. In addition, two classical bits are needed by Bob to announce Alice the outcome collection $\{(U_{i_1j_1}^A A_1, A_2),(B_1,U_{k_1l_1}^B B_2)\}$ belongs to through classical channel. Therefore, the information-theoretical efficiency of the proposed protocol is $\eta = (2+2)/(4+2) = 66.7\%$.

**(3) Comparisons with Gao's protocols**[36]

Here we compare the proposed protocol with Gao's protocols[36] in detail, since both of them belong to the kind of information leakage resistant quantum dialogue protocols using the entanglement swapping between Bell states. (a) Quantum resource. In every round communication, both the proposed protocol and Gao's protocols only need two Bell states, thus the same quantum resource is needed. (b) Quantum measurement. In every round communication, both the proposed protocol and Gao's protocols only need three Bell-basis measurements, thus the same number of quantum measurement is needed. (c) The information-theoretical efficiency. In Gao's protocols, Alice and Bob exchange their individual two bits per round communication by using four qubits and two bit classical information, thus their information-theoretical efficiency is $\eta = (2+2)/(4+2) = 66.7\%$. As a result, the information-theoretical efficiency of the proposed protocol is equal to its counterpart of Gao's protocols. Therefore, it can be concluded that the proposed protocol has the same performance as Gao's protocols in quantum resource, quantum measurement and the information-theoretical efficiency.

However, there are still big differences between them. We also take the first two Bell states $(A_1,B_1)$ and $(A_2,B_2)$ for an example. (a) The encoded Bell states are different. In the proposed protocol, the two initial Bell states are encoded by Alice and Bob, respectively. Concretely speaking, $A_1$ from $(A_1,B_1)$ is encoded by Alice with $U_{i_1j_1}^A$, while $B_2$ from $(A_2,B_2)$ is encoded by Bob with $U_{k_1l_1}^B$. However, in Gao's protocols, the same Bell state, derived from the entanglement swapping between two initial Bell states, is encoded by both Alice and Bob. Concretely speaking, $(B_1,B_2)$ (or $(A_1,A_2)$) is encoded by both Alice and Bob. (b) The key parameters needed for decoding are different. The proposed protocol needs two key parameters for decoding, i.e., the Bell-basis measurement of the initial shared secret Bell state and the corresponding relation between the two encoded initial Bell states and their entanglement swapping outcome. Concretely speaking, in order to know the prepared initial state of group 1, $(A_2,B_2)$ needs to be measured with Bell-basis by Bob. Moreover, the Bell-basis measurement outcomes of $(U_{i_1j_1}^A A_1, A_2)$ and $(B_1,U_{k_1l_1}^B B_2)$ are needed to infer the four possible kinds of initial states about $\{(U_{i_1j_1}^A A_1, B_1),(A_2,U_{k_1l_1}^B B_2)\}$ by both Alice and Bob. However, Gao's protocols only need one key parameter for decoding, i.e., the measurement correlation of two Bell states derived from the entanglement swapping between two initial Bell states. That is to say, in Gao's first protocol, in order to decode Bob's secret, Alice needs to use the Bell-basis measurement outcome of $(A_1,A_2)$ to infer its counterpart of $(B_1,B_2)$; in Gao's second protocol, in order to decode Alice's secret, Bob needs to use the Bell-basis measurement outcome of $(B_1,B_2)$ to infer its counterpart of $(A_1,A_2)$. (c) The methods for avoiding the information leakage problem are different. As analyzed above, in the proposed protocol, besides the entanglement swapping between any two Bell states, the shared secret Bell state also plays a key role in avoiding the information leakage problem. However, in Gao's protocols, the measurement correlation of two Bell states derived from the entanglement swapping between two initial Bell states helps overcome the information leakage problem.

Therefore, it can be concluded that compared with Gao's protocols, the proposed protocol provides another novel alternative idea for constructing an information leakage resistant quantum dialogue protocol using the entanglement swapping between Bell states.

## 5 Conclusions

To sum up, we have put forward a quantum dialogue protocol, which is based on the entanglement swapping between any two Bell states and the shared secret Bell state. The proposed protocol integrates the ideas of block transmission [4,7], two-step transmission[4,7] and unitary operation encoding[7] together using the Bell states as the information carriers. Besides the entanglement swapping between any two Bell states, a shared secret Bell state is also used to overcome the information leakage problem. The shared secret Bell state has two functions in the dialogue process: firstly, it makes Bob aware of the prepared initial state; secondly, it is used for Bob's encoding and entanglement swapping. Security analysis shows that the proposed protocol can resist the general active attacks from an outside eavesdropper Eve. Moreover, the relation between the maximal amount of information Eve can gain and the detection probability is derived.

**Acknowledgement**

The authors would like to thank the anonymous reviewers for valuable comments and suggestions on the improvement of this article. Funding by the National Natural Science Foundation of China (Grant No.11375152), and the Natural Science Foundation of Zhejiang Province (Grant No. LQ12F02012) is also gratefully acknowledged.